\documentclass[reprint,aps,pre,amsmath,amssymb,footinbib,floatfix]{revtex4-2}

\usepackage{adjustbox}
\usepackage{graphicx}
\graphicspath{{img/}}

\makeatletter
\renewcommand{\fnum@figure}{FIG. \thefigure}
\makeatother

\usepackage{amsfonts}
\usepackage{mathtools}
\usepackage[utf8]{inputenc}
\usepackage[T1]{fontenc}
\usepackage{t1enc}
\usepackage{tgtermes}

\usepackage[english]{babel}
\usepackage{afterpage}
\usepackage{xspace}

\usepackage[usenames,dvipsnames]{xcolor}

\usepackage[colorlinks,linktoc=all,pdfstartview=FitH,bookmarks=false]{hyperref}
\hypersetup{citecolor=MidnightBlue}
\hypersetup{linkcolor=MidnightBlue}
\hypersetup{urlcolor=MidnightBlue}

\newcommand{\bid}{\textbf{\textit{d}}} \newcommand{\bis}{\textbf{\textit{s}}}
\newcommand{\bbid}{\bar{\textbf{\textit{d}}}}
\newcommand{\bbis}{\bar{\textbf{\textit{s}}}}
\newcommand{\bit}{\textbf{\textit{t}}} \newcommand{\biB}{\textbf{\textit{B}}}
\newcommand{\mbzero}{\left<\bar{\pmb{\beta_0}}\right>}
\newcommand{\mbone}{\left<\bar{\pmb{\beta_1}}\right>}
\newcommand{\ccNP}{\mathsf{NP}}  
\newcommand{\ccP}{\mathsf{P}}  
\usepackage{circledsteps}
\pgfkeys{/csteps/inner ysep=2pt}
\pgfkeys{/csteps/inner xsep=2pt}

\usepackage{pdfpages} 
\usepackage{pgffor} 

\makeatletter
\AtBeginDocument{\let\LS@rot\@undefined}
\makeatother

\def\supplementfilename{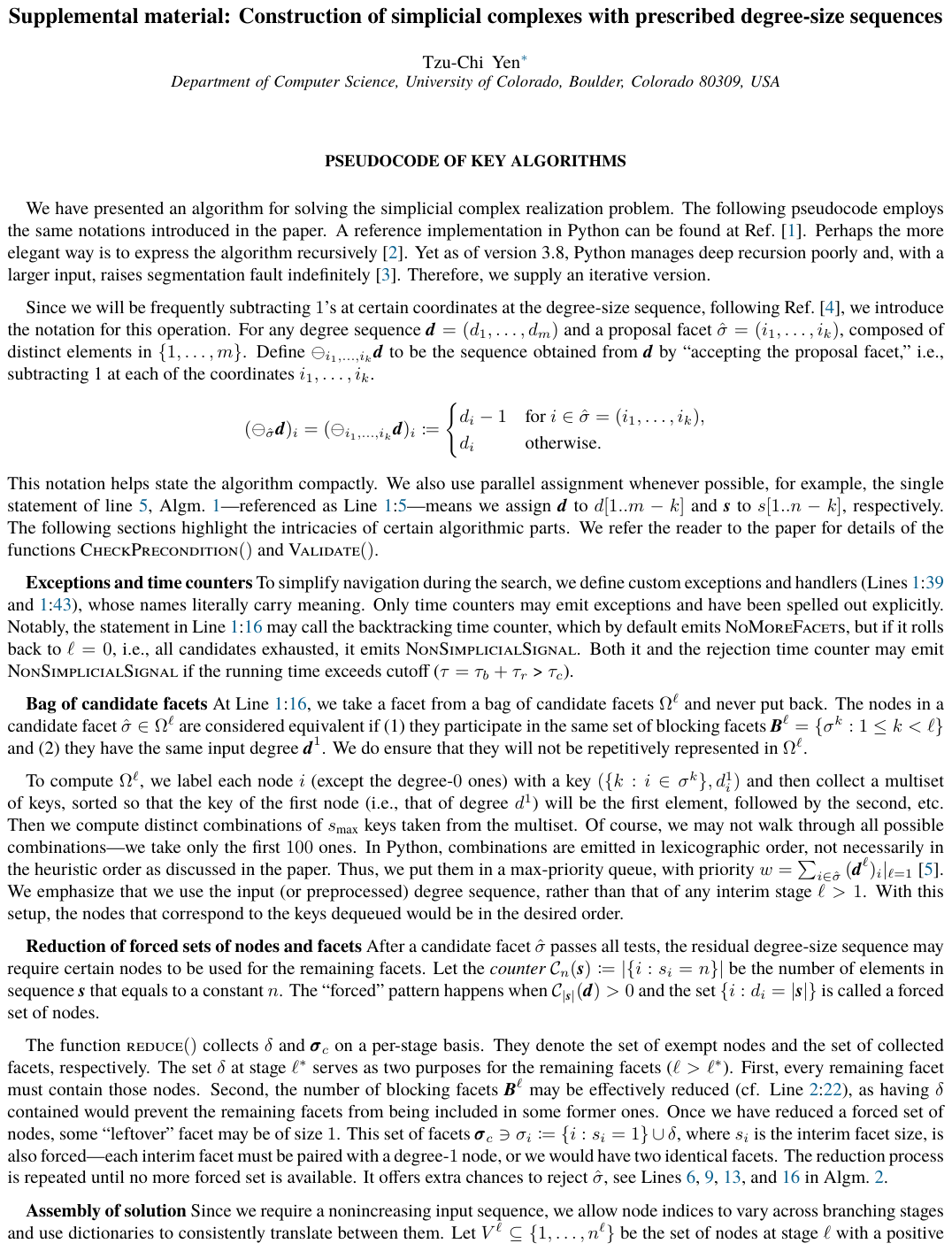}

\def\numbersupplementpages{\the\pdflastximagepages}

\newif\ifarXiv
\arXivtrue

\begin{document}

\title{Construction of simplicial complexes with prescribed degree-size sequences}

\author{Tzu-Chi Yen}
\email{tzuchi.yen@colorado.edu}
\affiliation{Department of Computer Science, University of Colorado, Boulder, Colorado 80309, USA}

\begin{abstract}
We study the realizability of simplicial complexes with a given pair of integer
sequences, representing the node degree distribution and the facet size
distribution, respectively. While the $s$-uniform variant of the problem is
$\mathsf{NP}$-complete when $s \geq 3$, we identify two populations of input
sequences, most of which can be solved in polynomial time using a recursive
algorithm that we contribute. Combining with a sampler for the simplicial
configuration model [J.-G. Young {\it et al.}, Phys. Rev. E {\bf 96}, 032312 (2017)],
we facilitate the efficient sampling of~simplicial ensembles from arbitrary degree
and size distributions. We find that, contrary to expectations based on dyadic
networks, increasing the nodes' degrees reduces the number of loops in simplicial
complexes. Our work unveils a fundamental constraint on the degree-size
sequences and sheds light on further analysis of higher-order phenomena based on
local structures.
\end{abstract}
\maketitle

Capturing higher-order dependencies is essential for the structural interpretation
of the organization and behavior of complex
systems~\cite{TBBE21,battistonNetworksPairwiseInteractions2020,bickWhatAreHigherorder2021}.
Simplicial complex modeling, among other methods in applied
topology~\cite{edelsbrunnerComputationalTopologyIntroduction2009,ghristElementaryAppliedTopology2014,otterRoadmapComputationPersistent2017},
provides a combinatorial description of the topology of the system, featuring
algebraic redundancies that may be compressed out using equivalence relations.
Similar to networks, simplicial complexes are composed of nodes that represent
system observables, and high-dimensional analogs of edges, called simplices,
that represent polyadic relationships among the nodes. Simplicial modeling made
several recent discoveries in complex systems, including the emergence of
discontinuous transitions~\cite{KB21} in epidemic
spreading~\cite{iacopiniSimplicialModelsSocial2019,landryEffectHeterogeneityHypergraph2020}
and synchronization~\cite{SA20,MTB20}, the multiscale hierarchy in adaptive
voter models~\cite{HK20}, the role of simplex size fluctuations in temporal
networks~\cite{petriSimplicialActivityDriven2018}, localization of
dynamics~\cite{st-ongeSocialConfinementMesoscopic2021a} and percolation
transitions~\cite{BZ18a,santosTopologicalPhaseTransitions2019,bobrowskiHomologicalPercolationEuler2020}.
In a related role, simplicial complexes have been used to summarize static
features, addressing questions about the local geometry of data, such as in
distinguishing the voting patterns in densely populated
cities~\cite{fengPersistentHomologyGeospatial2021} and in describing the shape
of scientific collaborations~\cite{pataniaShapeCollaborations2017}.

Dynamical processes on networks depend crucially on the network
structure~\cite{BBV08,PG16}. However, their generalizations to simplicial
structures and nonpairwise interactions are relatively less understood. Notably,
the basic question of which degree-size sequences can be realized by a complex
is still open. In this Letter we make progress in this direction by extensive
numerical experiments.

\begin{figure}[!b]
  \centering
  \begin{adjustbox}{center}
  \includegraphics[width=1\columnwidth]{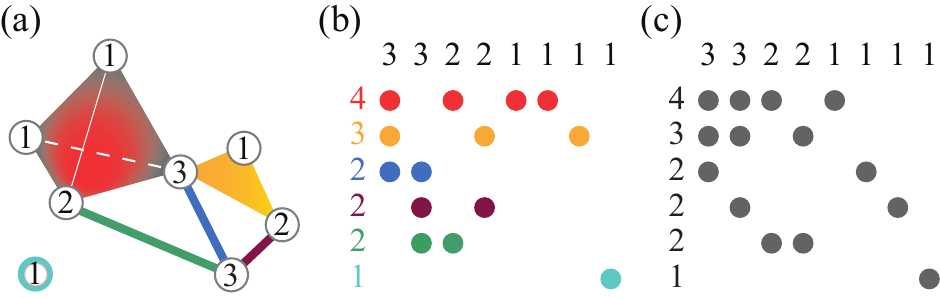}
  \end{adjustbox}
  \caption{(a) Geometric representation $|K|$ of a simplicial complex $K$ with
  degree-size sequence $\bid = \lbrace 3, 3, 2, 2, 1, 1, 1, 1 \rbrace$ and $\bis
  = \lbrace 4, 3, 2, 2, 2, 1 \rbrace$. The numbers on the nodes represent their
  degrees. (b) Its incidence matrix, where the circles represent $1$'s and each
  row (respectively column) constitutes a facet (respectively node). (c) An
  alternative realization of the same sequence, following the algorithm
  described in the main text.}
  \label{fig:intro}
\end{figure}

Let $V$ be a finite set of nodes. A simplicial complex on $V$ is a collection
$K$ of nonempty subsets of $V$, called simplices, subject to two requirements:
First, for each node $v \in V$, the singleton $\lbrace v \rbrace \in K$; second,
for all simplices $\tau \in K$, all its proper subsets $\sigma \subset \tau$
must also be in $K$. A {\it facet} is a maximal simplex, i.e., one that is not a
subset of any other simplex. Note that a simplicial complex can be fully
specified by listing only its facets, and we follow that convention here. We
define the degree $d_i$ of a node $v_i$ as the number of facets incident on
$v_i$ and the size (cardinality) $s_j$ of a facet $\sigma_j$ as the number of
nodes it contains. For any simplicial complex $K$, there is a corresponding
degree-size sequence $\bit_K=(\bid, \bis)$, where $\bid = \lbrace d_1, d_2,
\dots, d_{n} \rbrace$ and $\bis = \lbrace s_1, s_2, \dots, s_{m} \rbrace$ (see
Fig.~\ref{fig:intro}). However, the reverse statement is not always true. There
are sequences $\bit$ that cannot be realized by any simplicial complex. Hence,
inspired by Ref.~\cite{youngConstructionEfficientSampling2017}, we pose the {\it
simplicial complex realization} problem: Given integer sequences $\bit = (\bid,
\bis)$, does there exist a simplicial complex $K$ with that degree-size
sequence? When the answer is affirmative, we call the sequence simplicial.

The degree-size sequence reflects the local coupling patterns of the system.
Models that constrain these features can often be used as null models that
detect salient structures. In particular, the simplicial configuration model
(SCM)~\cite{youngConstructionEfficientSampling2017} specifies a distribution of
simplicial complexes with fixed degree-size sequences, and can be sampled via a
Markov chain Monte Carlo (MCMC) algorithm. The SCM extends the configuration
model for
graphs~\cite{bollobasProbabilisticProofAsymptotic1980,fosdickConfiguringRandomGraph2018}
and similar efforts in simplicial complexes of equal-size
facets~\cite{courtneyGeneralizedNetworkStructures2016}. Critically, the MCMC
requires an initial simplicial configuration to work, restricting its use to
empirical data which can act as the initialization. The algorithm we propose
yields an initialization for arbitrary simplicial sequences, not necessarily
taken from an observed data set. 

{\it Related work.}---A key difficulty in {\it simplicial complex realization}
is that no facet is allowed to be completely included in any other facet; we
call this the ``no-inclusion constraint.'' In contrast, hypergraphs have no such
constraint. For simple hypergraphs, Ref.~\cite{dyerSamplingHypergraphsGiven2020}
gives a rejection sampling algorithm that samples realizations with given
degrees and hyperedge sizes. For their non-simple counterparts,
Ref.~\cite{chodrowConfigurationModelsRandom2020} considers an MCMC approach for
generating such hypergraphs and
Ref.~\cite{arafatConstructionRandomGeneration2020} ensures the existence of a
starting realization which, however, needs not be simplicial. If we relax the
notion of facets and consider instead the degree sequence of nodes partaking in
given motifs, Refs.~\cite{KN10,MSMD21} use generating functions to
study such networks.

Testing whether a degree-size sequence is simplicial is a generalization of the
{\it graph realization} problem, in which the main result is the
Erd\H{o}s--Gallai theorem~\cite{paulerdosGraphsPrescribedDegrees1960} that
exactly characterizes graphical sequences with a set of easy-to-test
inequalities. Equivalently, a particular graph realization can be constructed by
a recursive application of the Havel--Hakimi
theorem~\cite{havelRemarkExistenceFinite1955,*hakimiRealizabilitySetIntegers1962}.
The reformulation of these theorems expedites the direct sampling of networks and
facilitates understanding of their properties (e.g., all scale-free networks are
sparse~\cite{delgenioAllScaleFreeNetworks2011}). Moreover, networks enjoy
tractable expressions for many ensemble-averaged quantities of interest (e.g.,
degree correlation, which tends to be disassortative in heterogeneous
networks~\cite{parkOriginDegreeCorrelations2003,PhysRevLett.104.108702}),
precisely because they are free from the no-inclusion constraint.
For simpliciality testing, however, none of these methods applies. 
Recently, Deza, Levin, Meesum, and
Onn~\cite{dezaOptimizationDegreeSequences2018} proved that deciding whether a
given sequence is the degree sequence of a simple $s$-uniform hypergraph is
$\ccNP$-complete when $s \geq 3$, through a reduction from the {\it 3-partition}
problem~\cite{gareyComputersIntractabilityGuide1979}. Interestingly, simple
$s$-uniform hypergraphs are equivalent to equidimensional simplicial complexes,
because when all hyperedges are the same size, they automatically satisfy the
no-inclusion constraint. This implies that deciding simpliciality {\it is} hard
and there may not exist an efficient algorithm to enumerate and sample these
instances. Yet, as our exploration reveals, not all sequence combinations are
equally hard. For example, for any $\bis$, taking the trivial degree sequence
$\bid= \lbrace 1, \dots, 1 \rbrace$ immediately yields a simplicial realization.

In this Letter, we develop a deterministic, backtracking-based search algorithm
that always correctly solves simpliciality, and present evidence that on most
instances it runs in polynomial time. We then explore the topology of the
constructed complex more generally via its Betti numbers, as well as using it as
a seed for the SCM ensemble. With the randomized realizations, we reveal the
regimes in which their expected topology changes as a function of the degree and
size sequences.

{\it Algorithm.}--- Our algorithm proceeds as follows. It is given as input a
node degree sequence~$\bid$ and a facet size sequence~$\bis$, where both
sequences are nonincreasing. Let $n = \lvert \bid \rvert$ and $m = \lvert \bis
\rvert$, where $| \cdot|$ stands for the cardinality. Simpliciality fails
trivially if there are fewer $1$'s in $\bid$ than in $\bis$, as it is doomed to
violate the no-inclusion constraint, or if $d_1 > m$ or $s_1 > n$, or
$\sum{\bid} \neq \sum{\bis}$, as it would be impossible to form an incidence
matrix. As a preprocessing step, we pair the $1$'s in $\bid$ with those in
$\bis$ to make a partial output.

Next, we pick $s_1$ nodes to make a candidate facet $\hat{\sigma}^{1}$. This
selection is not done stochastically, but favors higher-degree nodes---the
candidate facet $\hat{\sigma}$ with the largest sum of input degrees
$w(\hat{\sigma}) = \sum_{i \in \hat{\sigma}}{d_i}$ is preferred. We ensure that
$\hat{\sigma}$ is not included in any existing facet, or we proceed with the
next one in heuristic order. We will validate~$\hat{\sigma}^{1}$ with a series
of rules. If it fails any of them, we take the next candidate, until we accept
and advance to pick $s_2$ nodes for $\hat{\sigma}^{2}$. For
$\hat{\sigma}^{\ell}$ ($\ell \geq 2$), the facets with larger
$w(\hat{\sigma}^{\ell})$ still attain precedence.

With this overall structure in mind, we now express the algorithm recursively.
At each branching stage $\ell$, the input is a 3-tuple $(\bid^{\ell},
\bis^{\ell}, \biB^{\ell})$, where $\bid^{\ell}$ is the residual degree sequence
and $\bis^{\ell}$ the residual size sequence, denoting the marginal sums of the
incidence matrix that still need to be fulfilled. In addition, we have a
collection of ``blocking'' facets $\biB^{\ell} \coloneqq \lbrace \sigma^k :1
\leq k < \ell \rbrace$, where each accepted facet $\sigma^{k}$ plays a role in
the no-inclusion constraint. After accepting $\hat{\sigma}^{\ell}$, the output
is again a 3-tuple $(\bid^{\ell+1}, \bis^{\ell+1}, \biB^{\ell+1})$. The
algorithm returns a simplicial realization if $\sum{\bis^{\ell}} = 0$, or a
negative result when the entire search tree has been traversed.

To validate a candidate facet $\hat{\sigma}^{\ell}$, we assume that
$\hat{\sigma}^{\ell}$ is accepted and virtually move to the next branching stage
to obtain a number of intermediate data, which must obey the validation rules
before we actually branch. Precisely, we compute the 3-tuple $(\bid^{\ell+1},
\bis^{\ell+1}, \biB^{\ell+1})$ at stage~$\ell + 1$ and the set of
non-shielding~nodes $Q^\ell \coloneqq V^{\ell} \setminus \hat{\sigma}^\ell$,
where $V^{\ell}$ is the set of nodes at stage~$\ell$. For clarity, we drop all
superscripts in the following developments.

{\it Rule 1 (for $\bid$ and $\bis$)}~\footnote{This rule is similar to, but
stronger than, the Gale--Ryser characterization of bigraphic pairs of
sequences~\cite{gale_1957,*ryser_1957}, which would allow $s_\text{max} = \lvert
\bid \rvert$ for non-terminal facets, thus violating the no-inclusion
constraint.}. We ask that $d_\text{max} \leq \lvert\bis \rvert$, and that
$s_\text{max} \leq \lvert \bid \rvert$, where equality in the latter requirement
is only allowed when $\lvert \bis \rvert = 1$.

{\it Rule 2 (for $\bid$, $\bis$, and $Q$).} We require that~$\lvert \bis
\rvert~\leq~\sum_{i \in Q}{d_i}$, as every subsequent facet must contain at
least one non-shielding node (i.e., element of $Q$) in order to not be included
in~$\hat{\sigma}$.

If equality holds, each subsequent facet must take exactly one non-shielding
node. To secure its availability, we can thus further require that $s_\text{max}
- 1 \leq \lvert \bid \rvert - \lvert Q \rvert$.

{\it Rule 3 (for $\bid$ and $\biB$).} Let $V'$ be the set of nodes with positive
residual degree. We require that, for every blocking facet $\sigma \in \biB$,
$V'$ be not a subset of $\sigma$.

While Rule 1 is essentially the precondition, Rules~2--3 are meant to cut the
combinatorial explosion as much as possible, but not prohibit any realizable
sequence from being realized. If accepting a candidate facet leads to $d_i =
\lvert \bis \rvert$ for any node $i$, these nodes are required to be used for
the remaining facets---we invoke a subroutine to recursively consume those
``forced'' degrees. A Python implementation of the algorithm is freely
available~\cite{yenSimplicialtestPythonLibrary2021}. The pseudocode is given in
the Supplemental Material~\cite{sm}.

\begin{figure}[!t]
  \centering
  \begin{adjustbox}{center}
  \includegraphics[width=1\columnwidth]{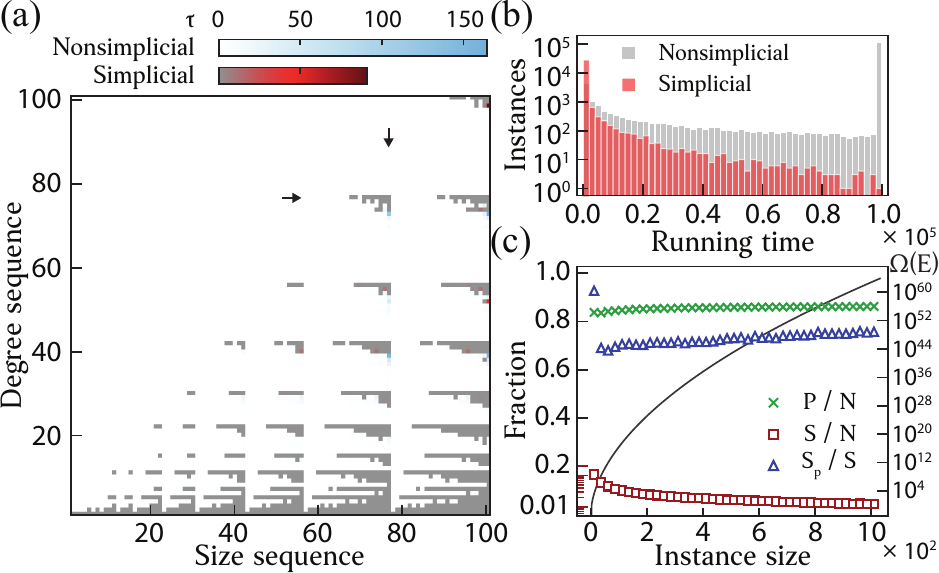}
  \end{adjustbox}
  \caption{(a) Realizability of all degree-size sequences with partitions of
  $13$, following ascending compositions. Color bars indicate running time
  $\tau$, where white to bluish colors mark the non-simplicial instances and
  gray to reddish colors mark the simplicial ones. Around $17$\% of the
  instances are hard. Qualitatively, more uniform degree sequences can pair with
  more size sequences to be simplicial and vice versa---see the arrows, which
  correspond to a sequence of $\lbrace3, 2, 2, 2, 2, 2\rbrace$. (b) Running time
  distribution when size sequence is fixed at $\bis = \lbrace 3, 3, \dots, 3
  \rbrace$ with $|\bis| = 20$, with $\bid$ set to all partitions of $60$. Around
  $84$\% of the instances are easy (not shown). (c) Simplicial ($\square$) and
  polynomial ($\times$; among all simplicial: $\triangle$) fractions versus
  instance size (bottom part in log scale) for $N=10^6$ uniform random
  partitions~\cite{nijenhuis_combinatorial_1978,thesagedevelopersSageMathSageMathematics2021}.
  The solid line shows the number of potential inputs at a specific $E$, i.e.,
  $\Omega(E) = a(E) \times a(E)$, where $a(n)$ is the number of partitions of
  $n$. In all cases, we apply a cutoff at $\tau = 10^5$.}
  \label{fig:landscape}
\end{figure}

{\it Easy \& hard instances.}---To benchmark the algorithm, we define the
running time as
\begin{equation*}
  \tau = \tau_{\text{b}} + \tau_{\text{r}} \ ,
\end{equation*}
where $\tau_{\text{b}}$ records the number of times the algorithm backtracks
(due to candidate depletion) and $\tau_{\text{r}}$ records the number of times a
proposed facet is rejected. These numbers are correlated: If we come up with
better validation rules, we reduce rejections and prevent backtracking. We call
a degree-size sequence easy (or polynomial) if $\tau = 0$, meaning that no
backtracking is necessary, the algorithm either finds a realization in
near-linear time or rejects simpliciality immediately. Otherwise, we call it
hard. Of course, this distinction is dependent on the choice of algorithm and
``hard'' instances with $\tau$ small will still be easy in practice, but for the
purposes of understanding the problem, we find it useful to distinguish between
those cases that are solved greedily by our algorithm and those that are not.

In Fig.~\ref{fig:landscape}(a), we show the realizability of all combinations of
degree-size sequences of a fixed instance size $E = \sum{\bid} = 13$. The
self-similar pattern reflects the recursive nature of the algorithm. However, we
are unable to conclude a recurrence relation in the spirit of Erd\H{o}s--Gallai,
due to the inherent complexity of solving particular instances. Indeed, there
are two distinct populations of easy and hard instances, where a major fraction
of inputs falls in the polynomial region. The easy majority can be understood as
the iterative application of the heuristic policy. For 3-uniform,
$\ccNP$-hard instances [Fig.~\ref{fig:landscape}(b)], most inputs are
polynomial and, on the average, the non-simplicial instances are harder than
simplicial ones, as tree exhaustion is needed. This highlights a useful property
that false negatives are kept at minimum when applying a reasonable cutoff.

To investigate the dependence of the simpliciality and hardness of degree-size
sequences on the instance size, we perform extensive numerical calculations,
generating uniform ensembles of sequences of random integers, $(\bid, \bis)$,
with a range up to $E=1010$. We test each sequence for simpliciality by applying
the algorithm and compute for each $E$ the simplicial fraction $S / N$, where
$S$ is the total number of simplicial sequences in the ensemble and $N$ is
the ensemble size, chosen at $N=10^6$. We also compute the fractions $P / N$ and
$S_p / S$, where $P$ is the total number of polynomial instances in the ensemble
and $S_p$ is the number of instances that are both simplicial and polynomial.
The results, plotted in Fig.~\ref{fig:landscape}(c), clearly demonstrate the
persistent existence of easy and hard populations at much larger sizes. An open
question is to what extent we can further separate these classes in polynomial
time, perhaps through better validation rules.

{\it Heuristic policy and topology}---When data are encoded as a simplicial
complex $K$, we can characterize their homotopical information by the Betti
numbers
$\beta_k(K)$~\cite{hatcherAlgebraicTopology2002,ghristElementaryAppliedTopology2014}.
They quantify the $k$-dimensional connectivity of objects by comparing their
volume and boundary at each dimension---$\beta_0$ is the number of connected
components, $\beta_1$ the number of homological cycles (or loops), while higher
$\beta_k$ effectively counts the number of $k$-dimensional cavities. The
topological cavities can be geometric in physical space, such as in granular
packings~\cite{saadatfarPoreConfigurationLandscape2017}, or abstract structures
in experimental
measurements~\cite{fengPersistentHomologyGeospatial2021,pataniaShapeCollaborations2017}.
For instance, a cavity in diffusion MRI readings could indicate axonal dropout,
a neurological disorder~\cite{sizemoreImportanceWholeTopological2018}. In the
following, we focus on the first two Betti numbers because they are easier to
interpret.

\begin{figure}[!b]
	\centering
  \begin{adjustbox}{center}
		\includegraphics[width=1\columnwidth]{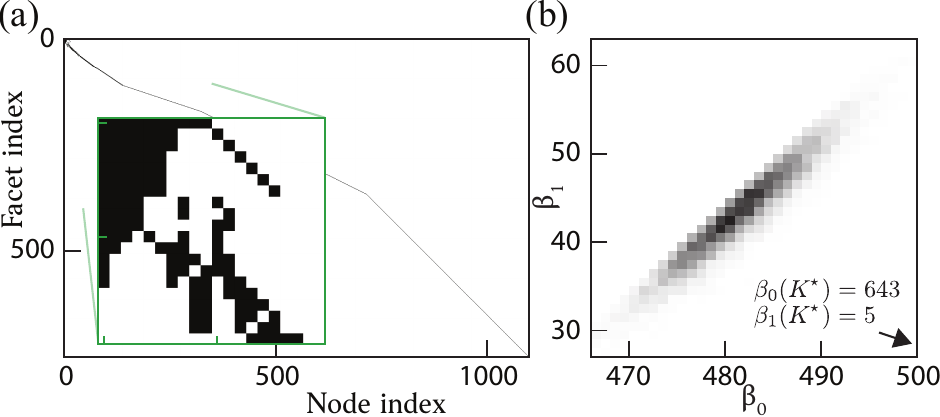}
	\end{adjustbox}
	\caption{(a) Realization $K^\star$ of the degree-size sequence from the human
	diseasome network (after pruning included
	faces)~\cite{youngConstructionEfficientSampling2017}, showing an assortative
	degree structure. Black squares show which nodes make which facet, with
	$n=1100$ and $m=752$. The indices are sorted such that nodes (respectively
	facets) with a larger degree (respectively size) have a lower index. The inset
	zooms in on the composition of the largest $20$ facets. (b) Joint Betti number
	distribution of $10^4$ randomized realizations of $K^\star$.}
  \label{fig:diseasome}
\end{figure}

An important feature of our heuristic policy is that high-degree nodes tend to
form larger facets, resulting in a core-periphery
structure~\cite{gallagherClarifiedTypologyCoreperiphery2021} with dangling and
isolated facets on the fringe. Therefore, the heuristic tends to find a
realization with a relatively large number of connected components and few
loops~\footnote{The latter case is due to an algorithmic choice to favor the
facets with small residual degrees, from among the candidates with the same
priority $w$~\cite{sm}.}. We show in Fig.~\ref{fig:diseasome}(a) this feature on
an empirical degree-size sequence extracted from the human diseasome
network~\cite{gohHumanDiseaseNetwork2007,youngConstructionEfficientSampling2017}.
Indeed, our algorithm can discover a realization with $\beta_0 = 643$ and
$\beta_1 = 5$ compatible~with~the degree-size sequence, whereas typical
realizations sampled from the SCM have much lower $\beta_0$ and much higher
$\beta_1$, see Fig.~\ref{fig:diseasome}(b). This algorithmic trait is consistent
across other datasets we examined~\cite{yenSimplicialtestPythonLibrary2021}.
More broadly, this priority policy shows a minimal example where adding
degree-size correlations can introduce atypical Betti numbers. This sheds light
on growth mechanisms that generate structures with specific homology, such as
being totally connected~\cite{horvatConnectednessMattersConstruction2020} or
containing many cycles~\cite{chujyoLoopEnhancementStrategy2021}.

\begin{figure}[!t]
  \centering
  \begin{adjustbox}{center}
  \includegraphics[width=1\columnwidth]{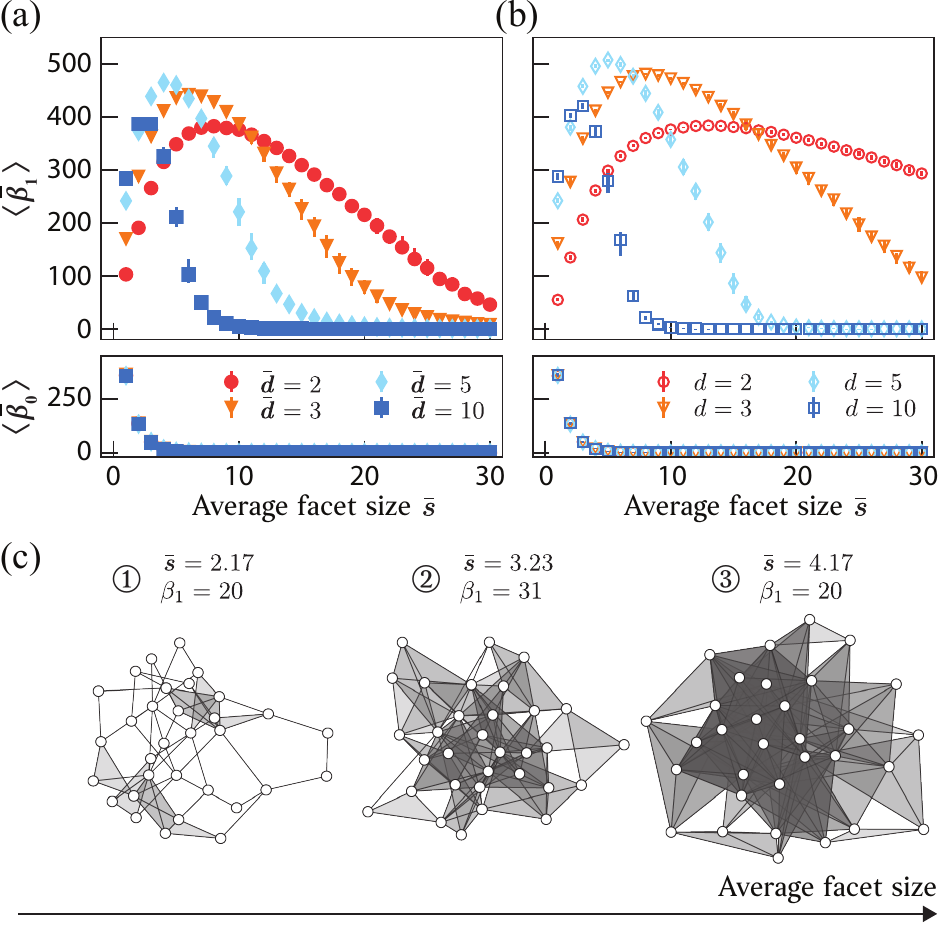}
  \end{adjustbox}
  \caption{(a) Average of the first two Betti numbers $\mbzero$ and~$\mbone$ of
  the simplicial complexes with Poisson-Poisson degree-size sequences, (b)
  Poisson size sequence and $d$-regular degree sequence. Each point shows the
  median of $10^2$ replicates of the indicated ensemble (see legend) and error
  bars show $25$\%--$75$\% quantiles. For each realization in the ensemble, we
  compute the average of Betti numbers from $10$ SCM replicates. All complexes
  have $E=10^3$. (c) Sketches of the simplicial structure. Enlarging the facets,
  while fixing the degree sequence, will first increase [\Circled{1} to
  \Circled{2}] and then decrease [\Circled{2} to \Circled{3}] the number of
  loops. The complexes have the same degree distribution at $\bbid = 2.86$.}
  \label{fig:synthetic}
\end{figure}

{\it SCM ensembles.}---To test the method, we generate random degree-size
sequences and test them for simpliciality. We generate from two Poisson
distributions, modified so that all facets of size $1$ are guaranteed to be
matched with a degree-$1$ node. For each simplicial sequence, we use the
constructed complex to initialize an MCMC sampler for the SCM
ensemble~\cite{youngConstructionEfficientSampling2017} and compute its mean
Betti numbers~$\bar{\pmb{\beta_0}}$ and~$\bar{\pmb{\beta_1}}$. Then we take the
average over the random partitions. Note that finding an initialization is
inefficient with a rejection sampler. The result, shown in
Fig.~\ref{fig:synthetic}, is a systematic study of~$\mbzero$ and~$\mbone$ with
respect to $\bbid$ and $\bbis$, where $\bbis$ is the mean facet size and $\bbid$
is the mean node degree, excluding the nodes that are created to match the
facets.

In Fig.~\ref{fig:synthetic}(a), we observe that the average number of connected
components $\mbzero$, in the SCM ensemble, decreases with increasing $\bbis$,
likely due to the reduction of facets with cardinality $1$ in the size sequence.
However, the distribution of cycles $\mbone$ is considerably more complicated.
In the low $\bbis$ regime, the simplicial complex acts as a dyadic network,
where denser networks contain more loops. By contrast, in the high $\bbis$
regime, the system is abundant in large simplices. Once there is a fraction of
higher-degree nodes, we have no other option but to bundle the large facets with
those nodes, resulting in a tree-like, few loops complex. We supply a parallel
analysis on $d$-regular degree distributions in Fig.~\ref{fig:synthetic}(b),
which tend to entail more loops than Poissonian ones with the same average
degree, as they possess fewer high-degree nodes.

The decay of the average number of cycles $\mbone$ when the average degree $\bbid$
is increased is reminiscent of the law of large numbers for Betti numbers in
random simplicial complexes (e.g., the Linial--Meshulam
model~\cite{linialPhaseTransitionRandom2016} or the random clique
complex~\cite{kanazawaLawLargeNumbers2021}). We note that in these studies, the
limiting behavior of Betti numbers is discussed in the context of increasing
facet density, where the decay is driven by filling the $k$-dimensional cycles
with $(k + 1)$-simplices. Here, the system has a fixed number of simplices and
the decay is driven by the no-inclusion constraints that prevent the realization
of any such complexes.

We also observe that $\mbone$ is unimodal with respect to mean facet size
$\bbis$ [Figs.~\ref{fig:synthetic}(a) and~\ref{fig:synthetic}(b)]. The unimodality comes from the
competitive relationship between the gradually multifaceted local geometry and
the depletion of available facets. When $\bbis$ is small, there are fewer large
facets to avoid for smaller ones and, thus, increasing the facet sizes has a
similar effect as densifying the degrees, which creates more loops [cf.
Fig.~\ref{fig:synthetic}(c), \Circled{1} to \Circled{2}]. That said, the loops
are destroyed as facets merge into larger ones [cf. Fig.~\ref{fig:synthetic}(c),
\Circled{2} to \Circled{3}]. This suggests the existence of an optimal facet
size distribution for loopy configurations, as the number of facets is limited.

{\it Discussion.}---In computational complexity, many graph problems are
$\ccNP$-hard in general, but may be in $\ccP$ for certain classes of
graphs~\cite{alekseevEasyHardHereditary2003}. {\it Simplicial complex
realization} is yet another addition to the list. We present a depth-bounded
branching algorithm whose complexity presents a rich structure. In particular,
simpliciality can be solved in time $f(m) E^{c}$ for some constant $c \approx
1$, where $E^{c}$ is the time spent in validations and the prefactor counts the
number of nodes in the execution tree. For easy instances, $f(m) = m$, and the
algorithm is linear in instance size. Otherwise, we have $f(m) = b^{m+1}$ in
general, where $b$ is the branching ratio that grows with instance size. We find
that $b$ is highly heterogeneous as a function of the branching stage---the
searcher stalls at mid-stages, not at the beginning or the end. It remains open
to accurately parametrize the complexity of hard instances of the simpliciality
problem, and to prove rigorously, if the algorithm runs in linear time on
average. Finally, we note that the branching design has opened up an avenue to
systematically improve the algorithm, for example, through stronger validation
rules to reduce the branching ratio, or introducing variants by
non-chronological backjumping or clause learning techniques, as critically used
in modern Boolean satisfiability (SAT)
solvers~\cite{marques-silvaConflictdrivenClauseLearning2009}. 

Aside from these computational complexity questions, the boundary between easy
and hard instances deserves further attention. We find that the instances tend
to be harder when $(\bid$, $\bis)$ contain numerous uniform entries, whereas a
Poisson-Poisson combination yields very little backtracking. The understanding
of when and why their hardness differs is poor compared to what is known for
constraint satisfaction problems~\cite{PhysRevLett.102.238701} or the inference
of stochastic block models~\cite{PhysRevE.84.066106}. This raises a number of
open questions, for example, is there an algorithmic phase transition that
separates easy and hard regions? Here, hard instances could mean either $\tau >
0$ or {\it really hard} in some other sense, e.g., takes exponential time, as
seen in SAT. Or, would there even be two different phase transitions? It is also
known that {\it graph isomorphism} can be solved in linear time for random
graphs~\cite{babaiCanonicalLabellingGraphs1979,babaiRandomGraphIsomorphism1980},
by leveraging the fact that in random graphs the degree distribution is
essentially never uniform, so that high-degree nodes help break symmetries. For
simpliciality, it could be useful to investigate the dependency of algorithmic
hardness on the degree sequence among equidimensional sequences. We believe that
the solutions to these problems will require new insights in the statistical
physics of
computation~\cite{zdeborovaStatisticalPhysicsInference2016,mooreNatureComputation2011},
notably, to identify the ``order parameter'' that characterizes the algorithmic
barrier to hard instances~\cite{cheesemanWhereReallyHard1991}.

In closing, we have developed a constructive algorithm to allow faster
realization of simplicial complexes with arbitrary degree-size sequences. Our
algorithm paves the way for a more comprehensive analysis of higher-order
phenomena in terms of local structural attributes, revealing their roles in
various dynamical systems.

{\it Acknowledgements.}---I am particularly grateful to Joshua A. Grochow for
significant discussions. I am grateful to Alice Patania and Jean-Gabriel Young
for helpful correspondence, as well as to Daniel B. Larremore for support. I
acknowledge the BioFrontiers Computing Core at the University of Colorado
Boulder for computing facilities. This work was funded in part by Grochow
startup funds.


\bibliography{simplicial}
\ifarXiv 

\section*{Supplementary material}

# Supplemental material: Construction of simplicial complexes with prescribed degree-size sequences

Tzu-Chi Yen[*]

*Department of Computer Science, University of Colorado, Boulder, Colorado 80309, USA*

## PSEUDOCODE OF KEY ALGORITHMS

We have presented an algorithm for solving the simplicial complex realization problem. The following pseudocode employs the same notations introduced in the paper. A reference implementation in Python can be found at Ref. [1]. Perhaps the more elegant way is to express the algorithm recursively [2]. Yet as of version 3.8, Python manages deep recursion poorly and, with a larger input, raises segmentation fault indefinitely [3]. Therefore, we supply an iterative version.

Since we will be frequently subtracting 1's at certain coordinates at the degree-size sequence, following Ref. [4], we introduce the notation for this operation. For any degree sequence $\boldsymbol{d} = (d_1, \ldots, d_m)$ and a proposal facet $\hat{\sigma} = (i_1, \ldots, i_k)$, composed of distinct elements in $\{1, \ldots, m\}$. Define $\ominus_{i_1, \ldots, i_k} \boldsymbol{d}$ to be the sequence obtained from $\boldsymbol{d}$ by "accepting the proposal facet," i.e., subtracting 1 at each of the coordinates $i_1, \ldots, i_k$.

$$(\ominus_{\hat{\sigma}} \boldsymbol{d})_i = (\ominus_{i_1, \ldots, i_k} \boldsymbol{d})_i := \begin{cases} d_i - 1 & \text{for } i \in \hat{\sigma} = (i_1, \ldots, i_k), \\ d_i & \text{otherwise.} \end{cases}$$

This notation helps state the algorithm compactly. We also use parallel assignment whenever possible, for example, the single statement of line 5, Algm. 1—referenced as Line 1:5—means we assign $\boldsymbol{d}$ to $d[1..m - k]$ and $\boldsymbol{s}$ to $s[1..n - k]$, respectively. The following sections highlight the intricacies of certain algorithmic parts. We refer the reader to the paper for details of the functions CheckPrecondition() and Validate().

**Exceptions and time counters** To simplify navigation during the search, we define custom exceptions and handlers (Lines 1:39 and 1:43), whose names literally carry meaning. Only time counters may emit exceptions and have been spelled out explicitly. Notably, the statement in Line 1:16 may call the backtracking time counter, which by default emits NoMoreFacets, but if it rolls back to $\ell = 0$, i.e., all candidates exhausted, it emits NonSimplicialSignal. Both it and the rejection time counter may emit NonSimplicialSignal if the running time exceeds cutoff ($\tau = \tau_b + \tau_r > \tau_c$).

**Bag of candidate facets** At Line 1:16, we take a facet from a bag of candidate facets $\Omega^\ell$ and never put back. The nodes in a candidate facet $\hat{\sigma} \in \Omega^\ell$ are considered equivalent if (1) they participate in the same set of blocking facets $\boldsymbol{B}^\ell = \{\sigma^k : 1 \le k < \ell\}$ and (2) they have the same input degree $\boldsymbol{d}^1$. We do ensure that they will not be repetitively represented in $\Omega^\ell$.

To compute $\Omega^\ell$, we label each node $i$ (except the degree-0 ones) with a key $(\{k : i \in \sigma^k\}, d_i^1)$ and then collect a multiset of keys, sorted so that the key of the first node (i.e., that of degree $d^1$) will be the first element, followed by the second, etc. Then we compute distinct combinations of $s_{\max}$ keys taken from the multiset. Of course, we may not walk through all possible combinations—we take only the first 100 ones. In Python, combinations are emitted in lexicographic order, not necessarily in the heuristic order as discussed in the paper. Thus, we put them in a max-priority queue, with priority $w = \sum_{i \in \hat{\sigma}} (\boldsymbol{d}^1)_i|_{\ell=1}$ [5]. We emphasize that we use the input (or preprocessed) degree sequence, rather than that of any interim stage $\ell > 1$. With this setup, the nodes that correspond to the keys dequeued would be in the desired order.

**Reduction of forced sets of nodes and facets** After a candidate facet $\hat{\sigma}$ passes all tests, the residual degree-size sequence may require certain nodes to be used for the remaining facets. Let the *counter* $\mathcal{C}_n(\boldsymbol{s}) := |\{i : s_i = n\}|$ be the number of elements in sequence $\boldsymbol{s}$ that equals to a constant $n$. The "forced" pattern happens when $\mathcal{C}_{|s|}(\boldsymbol{d}) > 0$ and the set $\{i : d_i = |\boldsymbol{s}|\}$ is called a forced set of nodes.

The function reduce() collects $\delta$ and $\boldsymbol{\sigma}_c$ on a per-stage basis. They denote the set of exempt nodes and the set of collected facets, respectively. The set $\delta$ at stage $\ell^*$ serves as two purposes for the remaining facets ($\ell > \ell^*$). First, every remaining facet must contain those nodes. Second, the number of blocking facets $\boldsymbol{B}^\ell$ may be effectively reduced (cf. Line 2:22), as having $\delta$ contained would prevent the remaining facets from being included in some former ones. Once we have reduced a forced set of nodes, some "leftover" facet may be of size 1. This set of facets $\boldsymbol{\sigma}_c \ni \sigma_i := \{i : s_i = 1\} \cup \delta$, where $s_i$ is the interim facet size, is also forced—each interim facet must be paired with a degree-1 node, or we would have two identical facets. The reduction process is repeated until no more forced set is available. It offers extra chances to reject $\hat{\sigma}$, see Lines 6, 9, 13, and 16 in Algm. 2.

**Assembly of solution** Since we require a nonincreasing input sequence, we allow node indices to vary across branching stages and use dictionaries to consistently translate between them. Let $V^\ell \subseteq \{1, \ldots, n^\ell\}$ be the set of nodes at stage $\ell$ with a positive

residual degree. Here residual means that we have accepted the candidate facet and (optionally) performed the reduction. We compute, between adjacent stages, a bijective map to specify the index of each node forward

$$f : V^\ell \to \{1, \dots, n^{\ell+1}\} \,,$$

such that the degree sequence for the next iteration is nonincreasing. In other words, for every pair of new indices $i, j \in \{1, \dots, n^{\ell+1}\}$, $i < j$ implies $d_i^{\ell+1} \geq d_j^{\ell+1}$. We use $f$ to transform the node indices in $\boldsymbol{B}^\ell$ forward; if a node has 0 residual degree, we simply remove its presence from $\boldsymbol{B}^{\ell+1}$. If the algorithm positively ended, we use the inverse map $f^{-1}$ to backward translate the node indices until $\ell = 1$ to create the solution, see the function Assemble().

---

**Algorithm 1** Pseudocode for solving the simplicial complex realization problem.

---

**Input:** A pair of nonincreasing arrays $\boldsymbol{d} = d[1..m]$ and $\boldsymbol{s} = s[1..n]$ of positive integers.
**Parameters:**
- $\tau_c \leftarrow 10^5$, cutoff, used implicitly in time counters.
- $\Xi_1, \Xi_2, \Xi_3 \leftarrow$ empty dictionaries, container for data at intermediate stages.

**Output:** A yes/no answer, and, if yes, the simplicial complex $K$ that realizes the degree-size sequence $(\boldsymbol{d}, \boldsymbol{s})$.

---

1: **function** IsSimplicial($\boldsymbol{d}, \boldsymbol{s}$)
2:      **if** CheckPrecondition($\boldsymbol{d}, \boldsymbol{s}$) = FALSE
3:          **return** FALSE, $\emptyset$
4:      Collect all facets of size 1 (if any). Let $k$ be the number of 1's in $s$. We have a partial output of $\bigcup_{i=0}^{k-1}\{\{m-i\}\}$.
5:      $\boldsymbol{d}, \boldsymbol{s} \leftarrow d[1..m-k], s[1..n-k]$
6:      **if** $\sum \boldsymbol{s} = 0$
7:          **return** TRUE, assemble($\{\}, 0$)
8:      $\boldsymbol{B}, \ell, \tau_b, \tau_r \leftarrow \{\}, 0, 0, 0$
9:      **while** simpliciality of $(\boldsymbol{d}, \boldsymbol{s})$ has not been decided
10:          $\ell \leftarrow \ell + 1$. Then $(\boldsymbol{d}^\ell, \boldsymbol{s}^\ell, \boldsymbol{B}^\ell) \leftarrow (\boldsymbol{d}, \boldsymbol{s}, \boldsymbol{B})$.
11:          Store current input variables, $\Xi_1[\ell] \leftarrow (\boldsymbol{d}^\ell, \boldsymbol{s}^\ell, \boldsymbol{B}^\ell)$
12:          $s_{\max} \leftarrow s[1]$. Then $\boldsymbol{s} \leftarrow s[2..n]$
13:          Compute the "bag" of candidate facets $\Omega^\ell$. Or, reuse the bag computed at a previous stage.
14:          **try**
15:              **while** no custom exception has been raised
16:                  Pick a facet $\hat{\sigma}$ of cardinality $s_{\max}$ from $\Omega^\ell$. Raise NoMoreFacets and increment $\tau_b$ by 1 if $|\Omega^\ell| = 0$.
17:                  **if** $\sum \boldsymbol{s} = 0$
18:                      **return** TRUE, assemble($\hat{\sigma}, \ell$)
19:                  $\hat{\boldsymbol{d}} \leftarrow \ominus_{\hat{\sigma}}\boldsymbol{d}$. Then compute the set of non-shielding nodes $Q$.        ⎫
20:                  **if** Validate($\hat{\boldsymbol{d}}, \boldsymbol{s}, \boldsymbol{B}, \hat{\sigma}, Q$) = FALSE            ⎬   Validating the candidate facet $\hat{\sigma}$.
21:                      $\tau_r \leftarrow \tau_r + 1$           ⎭   See paper for details.
22:                      **continue**
23:                  ind, $(\hat{\boldsymbol{d}}, \hat{\boldsymbol{s}}, \hat{\boldsymbol{B}}, \delta, \boldsymbol{\sigma}_c) \leftarrow$ reduce($\hat{\boldsymbol{d}}, \boldsymbol{s}, \boldsymbol{B} \cup \{\hat{\sigma}\}$)     ⎫
24:                  **if** ind = FALSE              ⎪
25:                      $\tau_r \leftarrow \tau_r + 1$        ⎬   Reducing the forced sets of nodes and facets.
26:                      **continue**             ⎪   See reduce.
27:                  **if** $\sum \hat{\boldsymbol{s}} = 0$            ⎪
28:                      **return** TRUE, assemble($\{\hat{\sigma}\} \cup \{\delta\} \cup \boldsymbol{\sigma}_c, \ell$)   ⎭
29:                  Compute $f$ and $f^{-1}$ using the positive coordinates of $\hat{\boldsymbol{d}}$.
30:                  $\Xi_3[\ell] \leftarrow (f, f^{-1})$                        ▷ See assemble.
31:                  $\hat{\boldsymbol{d}} \leftarrow \hat{\boldsymbol{d}}$, sorted in nonincreasing order            ⎫
32:                  $\hat{\boldsymbol{B}} \leftarrow \hat{\boldsymbol{B}}$, with node indices forward transformed using $f$    ⎪
33:                  **if** the tree branch $(\hat{\boldsymbol{d}}, \hat{\boldsymbol{s}}, \hat{\boldsymbol{B}})$ is explored       ⎪   We can make the algorithm faster by checking
34:                      $\tau_r \leftarrow \tau_r + 1$             ⎬   whether a branch is marked before we
35:                      **continue**             ⎪   recursively explore it.
36:                  Mark $(\hat{\boldsymbol{d}}, \hat{\boldsymbol{s}}, \hat{\boldsymbol{B}})$ as explored. Then $(\boldsymbol{d}, \boldsymbol{s}, \boldsymbol{B}) \leftarrow (\hat{\boldsymbol{d}}, \hat{\boldsymbol{s}}, \hat{\boldsymbol{B}})$.   ⎭
37:                  $\Xi_2[\ell] \leftarrow (\hat{\sigma}, \delta, \boldsymbol{\sigma}_c)$                        ▷ See assemble.
38:                  **break**
39:          **except** NoMoreFacets
40:              Rollback the state to a previous stage $\ell'$ by $(\boldsymbol{d}, \boldsymbol{s}, \boldsymbol{B}) \leftarrow \Xi_1[\ell']$     ⎫
41:              **delete** $\Omega^\ell$ ($\ell \geq \ell' + 1$)             ⎬   We call the process backtracking if $\ell' = \ell - 1$,
42:              $\ell \leftarrow \ell' - 1$             ⎭   or backjumping if $1 \leq \ell' < \ell - 1$.
43:          **except** NonSimplicialSignal
44:              **return** FALSE, $\emptyset$

---

**Algorithm 2** Pseudocode for the subroutines used in Algm. 1.

---

1: **function** REDUCE($d, s, B$)
2: $\quad \delta, \sigma_c \leftarrow \{\}, \{\}$
3: $\quad$ **while** $\mathcal{C}_{|s|}(d) > 0$ **and** $\sum s \neq 0$
4: $\qquad s_i \leftarrow s_i - \mathcal{C}_{|s|}(d)$ for all $i$
5: $\qquad$ **if** $(|s| > 1$ **and** $\mathcal{C}_0(s) > 0)$ **or** $\min s < 0$
6: $\qquad\quad$ **return** FALSE, $\emptyset$ $\qquad\qquad\qquad\qquad\qquad\qquad\qquad\qquad\qquad$ ▷ We use $\emptyset$ to denote a 5-tuple of repeated null's.
7: $\qquad \delta \leftarrow \delta \cup \{i : d_i = |s|\}$
8: $\qquad$ **if** $\sum s = 0$ **and** $\delta$ is a subset of some $\sigma \in B$
9: $\qquad\quad$ **return** FALSE, $\emptyset$
10: $\qquad d_i \leftarrow 0$ if $d_i = |s|$ for all $i$
11: $\qquad$ **if** $\mathcal{C}_1(s) > 0$
12: $\qquad\quad$ **if** $\mathcal{C}_1(s) > \mathcal{C}_1(d)$
13: $\qquad\qquad$ **return** FALSE, $\emptyset$
14: $\qquad\quad$ Find the nodes $i \subseteq \{i : d_i = 1\}$ to pair with the interim size-1 facets. $\qquad$ ▷ Use $B$ to satisfy the no-inclusion constraint.
15: $\qquad\quad$ **if** the number of such nodes $|i| < \mathcal{C}_1(s)$
16: $\qquad\qquad$ **return** FALSE, $\emptyset$
17: $\qquad\quad \sigma_c \leftarrow \sigma_c \cup \{i^* \cup \delta : i^* \in i[1..\mathcal{C}_1(s)]\}$
18: $\qquad\quad d \leftarrow \ominus_{i[1..\mathcal{C}_1(s)]} d$ $\qquad\qquad\qquad\qquad\qquad\qquad\qquad\qquad\qquad\qquad$ ▷ Set $\mathcal{C}_1(s)$ degree-1 coordinates to 0.
19: $\qquad\quad s \leftarrow s$, with all size-1 coordinates removed
20: $\qquad\quad$ **if** $\sum s = 0$
21: $\qquad\qquad \delta \leftarrow \{\}$
22: $\quad B \leftarrow \{\sigma \in B : \sigma \supseteq \delta\}$
23: $\quad$ **return** TRUE, $(d, s, B, \delta, \sigma_c)$
24:
25: **function** ASSEMBLE($\sigma, \ell$)
26: $\quad \sigma \leftarrow \sigma$, devoid of empty subsets
27: $\quad$ **if** $\ell > 1$
28: $\qquad$ **for** $\ell^* \leftarrow \ell - 1$ **down to** 1
29: $\qquad\quad f^{-1} = \Xi_3[\ell^*].f^{-1}$ $\qquad\qquad\qquad\qquad\qquad\qquad\qquad\qquad$ } Data $\Xi_2$ and $\Xi_3$ are implicitly required.
30: $\qquad\quad \mathcal{D} = \Xi_2[\ell^*]$ $\qquad\qquad\qquad\qquad\qquad\qquad\qquad\qquad\qquad\qquad$ } See Lines 1:30 and 1:37.
31: $\qquad\quad \sigma \leftarrow \sigma$, with node indices backward translated using $f^{-1}$
32: $\qquad\quad \sigma \leftarrow \{\sigma \cup \mathcal{D}.\delta : \sigma \in \sigma\}$
33: $\qquad\quad \sigma \leftarrow \sigma \cup \{\mathcal{D}.\hat\sigma\} \cup \mathcal{D}.\sigma_c$
34: $\quad \sigma \leftarrow \sigma \cup$ "the partial output collected at the preprocessing step" (Line 1:4)
35: $\quad$ **return** $\sigma$

---

* tzuchi.yen@colorado.edu

 
\fi

\end{document}